\documentclass[english,aip, numerical,jcp,reprint,twocolumn,floatfix]{revtex4-1}

\newcommand{\Ref}[1]{Ref.~\onlinecite{#1}}
\newcommand{\Refs}[1]{Refs.~\onlinecite{#1}}
\usepackage{graphicx}
\usepackage{dcolumn}
\usepackage{bm}
\usepackage{amsmath}
\usepackage{color}

\begin{document}

\preprint{AIP/123-QED}

\title{The Chemical Space of B,~N-substituted Polycyclic Aromatic Hydrocarbons: Combinatorial Enumeration and High-Throughput First-Principles Modeling}
\author{Sabyasachi Chakraborty}
\affiliation{Tata Institute of Fundamental Research, Centre for Interdisciplinary Sciences, Hyderabad 500107, India}

\author{Prakriti Kayastha}
\affiliation{Tata Institute of Fundamental Research, Centre for Interdisciplinary Sciences, Hyderabad 500107, India}

\author{Raghunathan Ramakrishnan}
\email{ramakrishnan@tifrh.res.in}
\affiliation{Tata Institute of Fundamental Research, Centre for Interdisciplinary Sciences, Hyderabad 500107, India}

\date{\today}

\begin{abstract}
Combinatorial introduction of heteroatoms in the two-dimensional framework of aromatic hydrocarbons opens up possibilities to design compound libraries exhibiting 
desirable photovoltaic and photochemical properties. 
Exhaustive enumeration and first-principles characterization of this chemical space provide indispensable insights for rational compound design strategies. 
Here, for the smallest seventy-seven Kekulean-benzenoid polycyclic systems, we reveal combinatorial substitution of C atom pairs with the isosteric and isoelectronic B, N pairs to result in 7,453,041,547,842 (7.4 tera) unique molecules. 
We present comprehensive frequency distributions of this chemical space, analyze trends and discuss a symmetry-controlled selectivity 
manifestable in synthesis product yield. Furthermore, by performing high-throughput {\it ab initio} density functional theory calculations of over thirty-three thousand (33k) representative molecules, we discuss quantitative trends in the structural stability and inter-property relationships across heteroarenes. 
Our results indicate a significant fraction of the 33k molecules to be electronically active in the 1.5--2.5 eV region, encompassing the most intense region of the solar spectrum, 
indicating their suitability as potential light-harvesting molecular components in photo-catalyzed solar cells.
\end{abstract}

\maketitle

%

\section{\label{sec:level1}Introduction}
Heteroarenes containing a subset of B, N, O, P, and S atoms are very versatile organic compounds exhibiting useful mechanical, optoelectronic, chemisorption and catalytic properties\cite{wong2006modulation,marcon2007tuning,campbell2010hydrogen,jiang2013heteroarenes,al2014water,hashimoto2014triplet,gong2015boron,stepien2016heterocyclic,ito2017annulative,wang2017ladder}. The most extensively studied heteroarenes are either those based 
on polycyclic aromatic hydrocarbon (PAH) molecules or their two-/three-dimensional (2D/3D) periodic forms: graphene and graphite. 
Borazine and hexagonal boron-nitride ($h-$BN) are fully heteroatom-substituted arenes; while their popular names, \emph{inorganic benzene} and \emph{inorganic graphite}, are inspired by their isoelectronicity with 
organic counterparts, their properties do exhibit stark contrast.
The latter compound, due to its suitable band structure properties, plays a vital role in the design of graphene-based heterostructures\cite{dean2010boron}; while in its partially hydrogenated form shows visible-light activity for photocatalyzed water splitting\cite{li2013semihydrogenated}. 
Both at the molecular level and in the extended 2D domain, B,~N-arenes exhibit unique physical and chemical properties\cite{dutta2008half,ci2010atomic,yamijala2013structural}. 
Due to the isoelectronic and isosteric relationships between C-C and B-N fragments, these compounds do exhibit similarities in chemistry to their parent hydrocarbon compounds--a fact that has motivated all the related synthesis endeavors during the past decades\cite{dewar1958624,dewar1959546,chissick1960new,dewar1964new,davies1967new,fang2006syntheses,jaska2006triphenylene,bosdet2007blue,yamamoto2007facile,bosdet2009bn,matsui2017one}. 
On the other hand, these heteroarenes have also been reported to exhibit mechanical and electronic properties differing from those of pure carbon and fully heteroatomic compounds\cite{miyamoto1994chiral,watanabe1996visible}.
Combinatorial diversity arising from site-specific atomistic substitutions in the arene framework combined with the local polarity 
introduced by heteroatoms gives rise to continuously distributed molecular properties in ranges desirable for a multitude of applications\cite{ghosh2011density,morgan2016efficient}. 
Yet another combinatorial scenario arises in extended arenes; for instance, nanotubules made of hexagonal BC$_2$N sheets show a wide degree of anisotropic conductivity stemming from distinct ways of rolling the heteroaromatic sheet into tubules\cite{miyamoto1994chiral}.  
From a molecular perspective, introduction of heteroatoms in the arene framework serves as an invaluable alternative to coupling C atoms with functional groups which in the case of aromatic arenes is thermodynamically amenable only at peripheral sites\cite{muller2014boron}.


Mathematical methods for enumerating molecular datasets have been thoroughly reviewed by Faulon\cite{faulon1992using}. Historically, isomer counting of cyclic compounds has been based on the enumeration theorem named after George P{\'o}lya\cite{polya1937kombinatorische,freudenstein1967basic,polya2012combinatorial}. Exhaustive applications of this technique have been carried out by Lindsey \textit{et al.} to enumerate macrocyclic compound libraries \cite{taniguchi2011virtual,taniguchi2013enumeration}suitable for light-harvesting\cite{yuen2018origin}. Baraldi \emph{et al.}\cite{baraldi2000regarding,baraldi1999cycle}have applied P{\'o}lya's theorem and enumerated the stereoisomers of highly symmetric icosahedral topologies. Graph theoretical methods came into light with comprehensive enumerations of polycyclic hydrocarbons through the pioneering works of Dias\cite{dias1985periodic,dias1986periodic,dias2007mathematics}, Balaban, \cite{balaban1985applications} and others\cite{lukvis2007growth}. Paton \emph{et al.} have enumerated saturated acyclic alkanes by explicitly accounting for the instability of strained carbon skeletons \cite{paton2007exploration}. Shao \emph{et al.} have applied subgroup decomposition of isomer permutation groups and enumerated a few fullerene cages\cite{shao1996enumeration2}and applied the same technique to list the isomers of B$_{24-m}$N$_m$\cite{shao1996enumeration1}. 

All the aforementioned works have been based on a  {\it non-constructive} strategy, {\it i.e.}, enumeration is based on closed-form algebraic expressions rather than explicit generation of molecular structures. As far as benzenoid hydrocarbons are concerned, a {\it constructive} strategy has been found to be of larger applicability, wherein explicit generation of structures or the corresponding 
graphs is required. For example, benzenoid hydrocarbons have been enumerated using the perimeter or the area of self-avoiding polygons on a hexagonal lattice\cite{gutman1989introduction,cyvin1992enumeration,voge2002number}. 
Using this approach the benzenoid compounds with up to 50 hexagonal cells have been enumerated with a parallel 
algorithm\cite{jensen2009parallel}. 
To date, one of the fastest approaches to constructively enumerate Kekulean benzenoid compounds---which is of interest to chemists---has been that of Brinkmann {\it et al.}\cite{brinkmann2007fusenes} which utilizes the idea of dual-graphs\cite{brinkmann2002constructive}. 
A detailed account of the Kekulean structure as an important descriptor for benzenoid hydrocarbons can be found in the work of Cyvin \emph{et al.}\cite{cyvin2013kekule}.

B,~N-arenes have been the subject of a diverse range of theoretical as well as combined theoretical/experimental studies. To begin with, already in the late 60's Hoffmann \emph{et al.} had applied the extended-H\"uckel molecular orbital theory to a dozen or so heteroaromatic compounds, discussed the role of intramolecular non-bonded electrostatic interactions in these compounds and commented on their stability \cite{hoffmann1964extended}. Prior to that, Dewar had reported the synthesis of hetero-phenanthrene with 9:10 C atoms replaced by B, N pairs \cite{dewar1958624}. Eventually, a number of investigations have studied larger compounds ranging from PAHs\cite{neue2013bn,wang2014straightforward,long2016doping,sanyal2014bn,wang2015b2n2,ishibashi2017bn,fias2018alchemical} to fullerene cages\cite{zhu1997alternant,jensen1993stability,zhu1995bn,seifert1997boron,evangelisti1997ab,pattanayak2002boron,balawender2018exploring} partly or completely enriched by B, N pairs. 
\begin{figure}[htbp!]
\centering
\includegraphics[width=8cm]{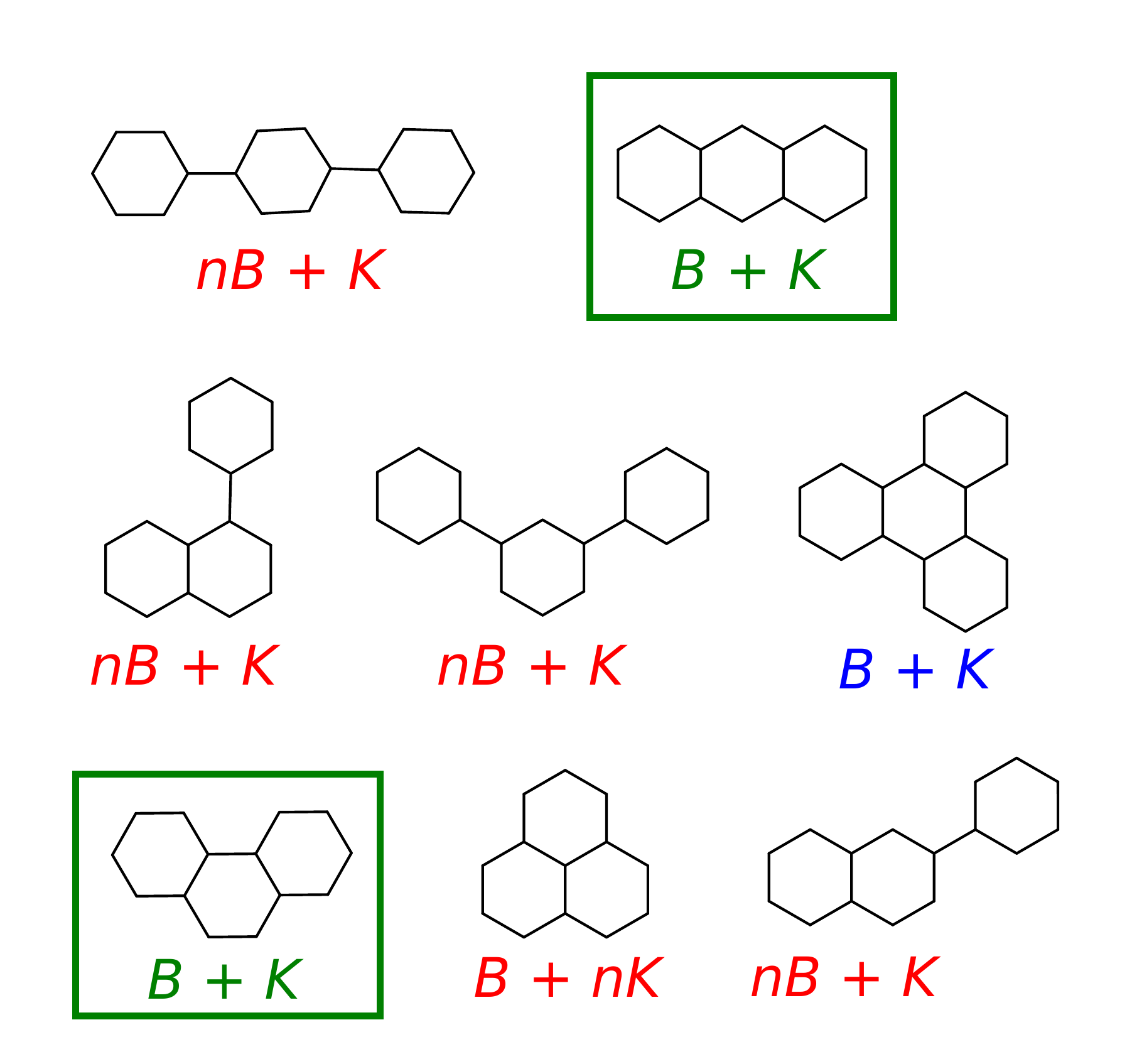} 
\caption{Generation of polycyclic aromatic hydrocarbon frameworks by connecting and  fusing three hexagons. {\it B}, {\it K}, {\it nB}, and {\it nK} stand for benzenoid,
Kekul{\'e}, non-benzenoid, and non-Kekul{\'e} frameworks, respectively.}
\label{fig01}
\end{figure}

The purpose of this paper is to non-constructively enumerate all possible unique compounds formed by substituting pairs of C atoms in PAHs comprising up to six cycles with B and N atoms. We utilize the nuclear permutation groups that are isomorphic to the rotation groups of the PAH scaffolds and generate the corresponding {\it pattern inventory}. One of the aims of this study is to provide consolidated tabulations of B,~N-substituted PAH (BN-PAH) compound frequencies---as a function of stoichiometries, symmetry and sites---to enable identification of statistical and combinatorial trends facilitating chemical space design strategies. As exemplars, we discuss (i) deviation of substitution patterns from commonly expected binomial/multinomial distributions; (ii) in the case of two or more PAHs 
of similar size ({\it i.e.} made of the same number of benzene rings) and symmetry (same point group), non-trivial frequency selectivities giving rise to distinct product distributions. We clarify the origin of both these effects using nuclear permutation groups. 
Furthermore, for a subset of 33,059 BN-PAH compounds, we present accurate geometrical features, energetics, optoelectronic properties, and inter-property correlations based on high-throughput density functional theory (DFT) calculations. %

\begin{figure*}[hpbt!]
\centering
\includegraphics[width=14.5cm]{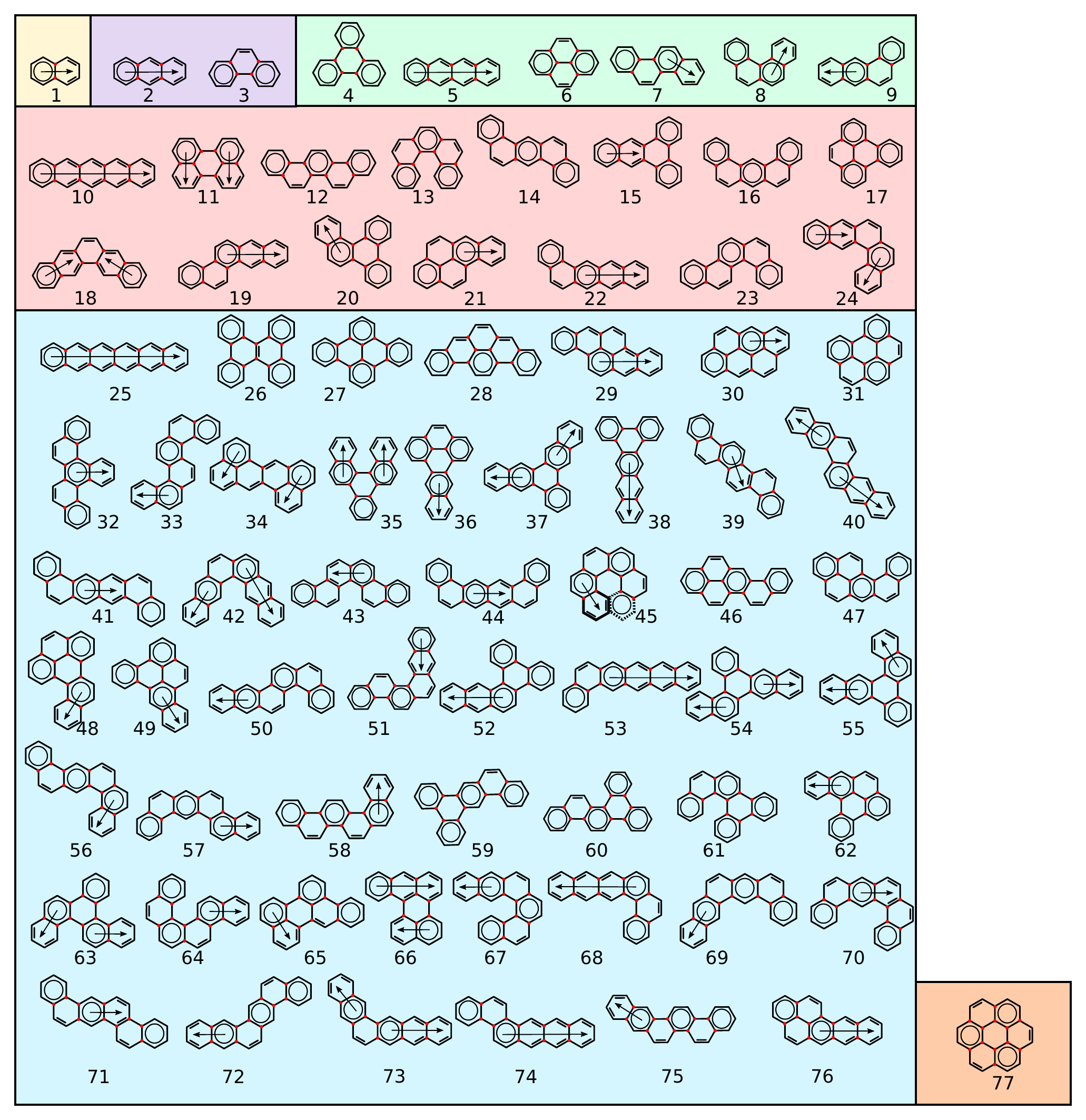} 
\caption{
Clar structures of the seventy-seven poly aromatic hydrocarbons comprising up to six benzene rings. Coronene, 
a formally seven-membered system, is shown as a special case (see discussions for more details): 
1)Naphthalene, 2)Anthracene, 3)Phenanthrene, 4)Isochrysene, 5)Tetracene, 6)Pyrene, 7)Chrysene, 8)Tetrahelicene, 9)Tetraphene, 10)Pentacene, 11)Perylene, 12)Picene, 13)Pentahelicene, 14)Benzo[k]tetraphene, 15)Benzo[f]tetraphene, 16)Benzo[m]tetraphene, 17)Benzo[e]pyrene, 18)Pentaphene, 19)Benzo[c]tetraphene, 20)Benzo[g]chrysene, 21)Benzo[a]pyrene, 22)Benzo[a]tetracene, 23)Benzo[c]chrysene, 24)Benzo[a]tetraphene, 25)Hexacene, 26)Dibenzo[g,p]chrysene, 27)Dibenzo[fg,op]tetracene, 28)Benzo[rst]pentaphene, 29)Dibenzo[c,pqr]tetraphene, 30)Dibenzo[def,mno]chrysene, 31)1,12-benzoperylene, 32)Benzo[s]picene, 33)Dibenzo[c,l]chrysene, 34)Dibenzo[de,mn]tetracene, 35)Naphtho[1,2-g]chrysene, 36)Dibenzo[de,qr]tetracene, 37)Benzo[h]pentaphene, 38)Dibenzo[a,c]tetracene, 39)Benzo[c]picene, 40)Naphtho[2,3-c]tetraphene, 41)Dibenzo[a,j]tetracene, 42)Naphtho[2,3-a]tetraphene, 43)Naphtho[1,2-c]chrysene, 44)Dibenzo[a,l]tetracene, 45) Hexahelicene, 46)Benzo[pqr]picene, 47)Dibenzo[m,pqr]tetraphene, 48)Dibenzo[ij,no]tetraphene, 49)Dibenzo[f,pqr]tetraphene, 50)Naphtho[2,1-c]tetraphene, 51)Naphtho[2,3-c]chrysene, 52)Dibenzo[a,c]tetraphene, 53)Benzo[a]pentacene, 54)Dibenzo[c,f]tetraphene, 55)Dibenzo[a,f]tetraphene, 56)Dibenzo[a,k]tetraphene, 57)Dibenzo[c,m]tetraphene, 58)Benzo[a]picene, 59)Dibenzo[f,k]tetraphene, 60)Benzo[f]picene, 
61)Benzo[f]pentahelicene, 62)Naphtho[1,2,3,4-pqr]tetraphene, 63)Dibenzo[c,p]chrysene, 64)Benzo[c]pentahelicene, 65)Benzo[b]perylene, 66)Benzo[a]perylene, 67)Benzo[b]pentahelicene,
68)Naphtho[1,2-a]tetracene, 69)Benzo[a]pentaphene, 70)Dibenzo[a,m]tetraphene, 71)Dibenzo[c,k]tetraphene, 72)Benzo[c]pentaphene, 73)Hexaphene, 74)Naphtho[2,1-a]tetracene, 
75)Benzo[b]picene, 76)Naphtho[2,1,8-qra]tetracene \& 77)Coronene. Names of PAHs are based on \Refs{NIST,sander1997polycyclic,ChemSpider}. For all molecules, inner-sites are marked with red dots.} 
\label{PAH}
\end{figure*}
\begin{figure*}[htbp!]
\centering
\includegraphics[width=12cm]{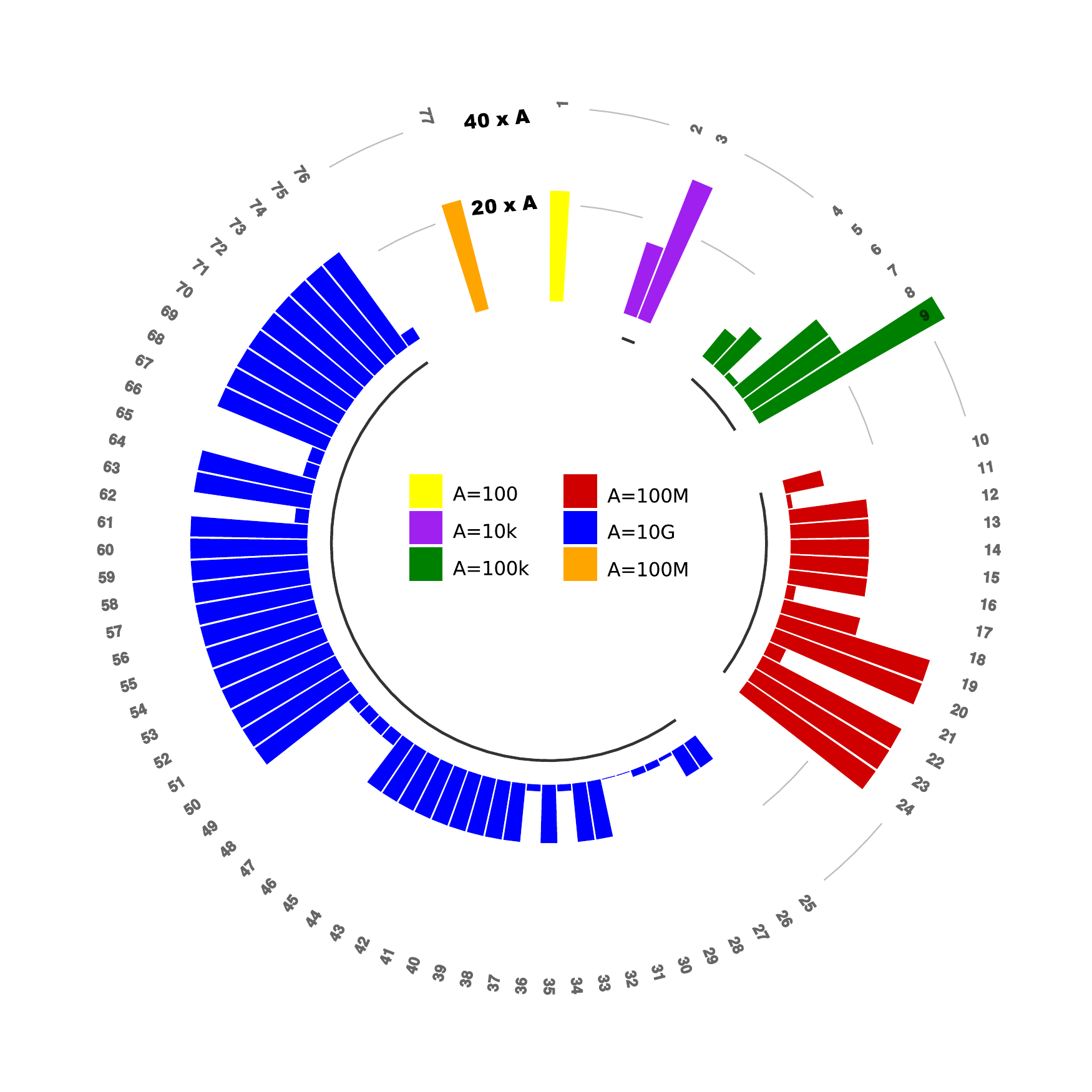}
\caption{Circular histogram depicting frequencies of all possible B,~N-substituted compounds of all seventy-seven PAHs. See Table \ref{tab:allcarbons} for more details.}
\label{fig:circular}
\end{figure*}
\begin{table*}[htbp!]
\caption{Number of constitutional isomers formed by substituting pairs of inner C atoms with B and N atoms. For every parent hydrocarbon, number of products are collected for various stoichiometries along with their total: $n$ is the number of rings in the parent hydrocarbon, $I$ is the index of the hydrocarbon in Fig.~\ref{PAH}, $X$ is the number of inner C atoms in the parent hydrocarbon available for substitution, and $\lbrace Z \rbrace$ lists the the cycle indices of the generator subgroup. Also given for every parent hydrocarbon, is the isomorphic point group of the molecular symmetry group ($\mathcal{G}^{\rm iso.}$).}
\centering
        \begin{tabular}{l l l  r r r r   r r r r  r} 
        \hline
$n$  & $I$ & $\mathcal{G}^{\rm iso.}$ & $X$ & C$_{X-2}$  & C$_{X-4}$& C$_{X-6}$& C$_{X-8}$& C$_{X-10}$ & C$_{X-12}$&  Total\textsuperscript{a} & $\lbrace Z \rbrace$\\
     &     &      &     &  B$_1$N$_1$  &  B$_2$N$_2$&  B$_3$N$_3$&  B$_4$N$_4$&  B$_5$N$_5$  &  B$_6$N$_6$&  & \\
\hline
 2 & 1 & $C_2$   & 2      & 1     &     & & & &   &  1 & \{$P_2^1$\}\\
 3 & 2 & $D_2$   & 4& 3   & 3     &     & & &   &  6 & \{$2P_2^2$\}\\
 3 & 3 & $C_2$   & 4& 6   & 4     &     & & &   &  10& \{$P_2^2$\}\\
 4 & 4 & $D_3$   & 6& 5   & 18    & 4   & & &   &  27& \{$P_3^2, P_2^3$\}\\
 4 & 5,6 & $D_2$ & 6& 8   & 27    & 6   & & &   &  82 & \{$P_2^3, P_2^2$\}\\
 4 & 7 & $C_2$   & 6& 15  & 48    & 10  & & &    &  73 & \{$P_2^3$\}\\
 4 & 8 & $C_2$   & 6& 16  & 48    & 12  & & &    &  76 & \{$P_2^2$\} \\
 4 & 9 & $C_1$   & 6& 30  & 90    & 20  & & &   &   140 & \{$P_1^6$\}\\
 5 & 10 & $D_2$  & 8& 14  & 114   & 140 &22 & &   &  290 & \{$2P_2^4$\}\\
 5 & 11 & $D_2$  & 8& 17  & 119   & 150 &24 & &   &  310 & \{$P_2^4,P_2^2$\} \\
 5 & 12--18      & $C_2$  & 8& 28 & 216 & 280&38 & &   &  3,934  & \{$P_2^4$ \} \\
 5 & 19--24      & $C_1$  & 8& 56 & 420 & 560  &70    &      & & 6,636 & \{$P_1^8$\}\\
 6 & 25--27      & $D_2$  &10& 23 & 330 & 1056 & 810 & 66 &  &  6,855 & \{$P_2^5,P_2^4$\} \\  
 6 & 28--41      & $C_2$  &10& 45 & 640 & 2100 & 1590 & 126 &  & 63,014 & \{$P_2^5$\} \\  
 6 & 42--45      & $C_2$  &10& 46 & 640 & 2112 & 1590 & 132 &  &  18,080 & \{$P_2^4$\} \\  
 6 & 46--76      & $C_1$  &10& 90 &1,260& 4,200 & 3,150 & 252 &  & 277,512 & \{$P_1^{10}$\}\\
 7 & 77          & $D_6$  &12& 14 & 274 & 1,586 & 2,976 & 1428 & 96 &  6,374 & \{$P_6^2, P_2^6, P_2^4$\} \\  
\hline
\end{tabular}
 \label{tab:innercarbons}
 \newline
\textsuperscript{a} For a given $n$, values are reported as sum over all $I$.
\end{table*}
\begin{table*}[htbp!]
\caption{
Number of constitutional isomers formed by substituting pairs of 
peripheral C atoms with B and N atoms. For every parent hydrocarbon, number of products are collected for various stoichiometries along with their total: $n$ is the number of rings in the parent hydrocarbon, $I$ is the index of the hydrocarbon in Fig.~\ref{PAH}, $X$ is the number of peripheral C atoms in the parent hydrocarbon available for substitution, and $\lbrace Z \rbrace$ lists the the cycle indices of the generator subgroup. Also given for every parent hydrocarbon, is the isomorphic point group of the molecular symmetry group ($\mathcal{G}^{\rm iso.}$).
}
        \begin{tabular}{l l l   r r r    r r r   r r r   r r  l } 
        \hline
$n$  & $I$ & $\mathcal{G}^{\rm iso.}$  & $X$ & C$_{X-2}$    & C$_{X-4}$  & C$_{X-6}$  & C$_{X-8}$  & C$_{X-10}$   & C$_{X-12}$ & C$_{X-14}$  & C$_{X-16}$  &Total\textsuperscript{a} & $\left\lbrace Z\right\rbrace$\\
     &     &      &     &  B$_1$N$_1$  &  B$_2$N$_2$&  B$_3$N$_3$&  B$_4$N$_4$&  B$_5$N$_5$  &  B$_6$N$_6$&  B$_7$N$_7$ & B$_8$N$_8$ & &  $\left\lbrace p_x^y,\,p_n^m p_l^k \right\rbrace$\\
\hline
 2 &  1 & $D_2$      & 8  & 14 & 114 & 140  &  22  &      &     &  &  &      290  & \{$2P_2^4$\}\\
 3 &  2 & $D_2$      & 10 & 23 & 330 & 1,056 &  810 &  66  &     &  &  &    2285  & \{$P_2^5,P_2^4$\} \\
 3 &  3 & $C_2$      & 10 & 45 & 640 & 2,100 & 1,590 & 126  &     &  &  &   4501  & \{$P_2^5$\} \\
 4 &  4 & $D_3$      & 12 & 22 & 510 & 3,084 & 5,820 & 2,772 & 166 &  & & 12,374  & \{$P_3^4, 2P_2^6$\} \\
 4 &  5 & $D_2$      & 12 & 33 & 765 & 4,620 & 8,730 & 4,158 & 246 &  & & 18,552  & \{$2P_2^6$\} \\
 4 &  6 & $D_2$      & 10 & 23 & 330 & 1,056 & 810 & 66 &  &  &  &    2,285  & \{$P_2^5, P_2^4$\} \\
 4 & 7,8 & $C_2$     & 12  & 66 & 1,500 & 9,240 & 17,370 & 8,316 & 472   &  &  &  73,928 & \{$P_2^6$\}\\
 4 &  9 & $C_1$      & 12 & 132 & 2,970 & 18,480 & 34,650 & 16,632 & 924   &  &  & 73,788 & \{$P_1^12$\} \\
 5 & 10 & $D_2$      & 14 & 46 & 1,533 & 15,030 & 52,710 & 63,108 & 21,126 & 868   &  & 154,421 & \{$P_2^7, P_2^6$\} \\
 5 & 11 & $D_2$      & 12 & 33 &  765 &  4,620 & 8,730 & 4,158 & 246 &  &  &  18,552  & \{$2P_2^6$\} \\
 5 & 12--15, 18 & $C_2$  & 14 & 91 & 3,024 & 30,030 & 105,210 & 126,126 & 42,112 & 1,716   &  &  1,541,545 & \{$P_2^7$\} \\
 5 & 16 & $C_2$   & 14 & 92 & 3,024 & 30,060 & 105,210 & 126,216 & 42,112 & 1,736   &  &   308,450 & \{$P_2^6$\} \\
 5 & 17 & $C_2$      & 12 & 66 & 1,500 & 9,240 & 17,370 & 8,316 & 472 &  &  &  73,928  & \{$P_2^6$\} \\
 5 & 19,20\& & $C_1$  & 14 & 182 & 6,006 & 60,060 & 210,210 & 252,252 & 84,084 & 3,432   &  &  3,081,130 & \{$P_1^14$\} \\
   & 22--24    & & &  & &  &  &  &  &    &  &    \\
 5 & 21       & $C_1$  & 12 & 132 & 2,970 & 18,480 & 34,650 & 16,632 & 924 & &  &  73,788 & \{$P_1^12$\} \\
 6 & 25,26    & $D_2$  & 16 & 60  & 2,772 & 40,040 & 225,540 & 504,504 & 420,840 & 102,960 & 3,270 & 2,599,972 & \{$2P_2^8$\} \\
 6 & 27       & $D_2$  & 14 & 46 & 1,533 & 15,030 & 52,710 & 63,108 & 21,126 & 868 &  &  154,421 & \{$P_2^7,P_2^6$\} \\
 6 & 28,29 \&  & $C_2$  & 14 & 91 & 3,024 & 30,030 & 105,210 & 126,126 & 42,112 & 1,716 &   & 1,233,236 & \{$P_2^7$\} \\
   & 34,36     &  &  &  &  &  &  &  &  &  &   &  & \\
 6 & 30,31           & $C_2$  & 12 & 66 & 1,500 & 9,240 & 17,370 & 8,316 & 472 &  &  &  73,928  & \{$P_2^6$\} \\
 6 & 32,33 \& & $C_2$  & 16 & 120 & 5,488 & 80,080 & 450,660 & 1,009,008 & 841,120 & 205,920 & 6,470 &  31,186,392 & \{$P_2^8$\} \\
   & 35,37--45 &   &  &  &  &  &  &  &  &  &  &   \\
 6 & 46--49,62 \&    & $C_1$  & 14 & 182 & 6,006 & 60,060 & 210,210 & 252,252 & 84,084 & 3,432 &  & 4,929,808 & \{$P_1^14$\} \\
   & 65,66,76    &   &  &  &  &  &  &  &  &  &  &   \\
 6 & 50--61,63 \& & $C_1$  & 16 & 240 & 10,920 & 160,160 & 900,900 & 2,018,016 & 1,681,680 & 411,840 & 12,870 & 119,522,398 & \{$P_1^{16}$\} \\
   & 64,67--75 &  &  &  &  &  &  &  &  &  &  &  \\
 7 & 77       & $D_6$  & 12 & 11  & 265 & 1,542 & 2,940 & 1,386 & 90 &  &  &  6,234   & \{$P_6^2, 2P_2^6$\} \\
\hline
\end{tabular}
\label{tab:PeripheralCarbons}
\newline
\textsuperscript{a} For a given $n$, values are reported as sum over all $I$.
\end{table*}

\begin{turnpage}
\begin{table*}[htbp!]
\caption{
Number of constitutional isomers formed by substituting pairs of 
all C atoms with B and N atoms. 
For every parent hydrocarbon, number of products are collected for various stoichiometries
along with their total: $n$ is the number of rings in the parent hydrocarbon, 
$I$ is the index of the hydrocarbon in Fig.~\ref{PAH}, and
$X$ is the number of C atoms in the parent hydrocarbon 
available for substitution. 
Also given for every parent hydrocarbon,
is the isomorphic point group of the molecular symmetry group ($\mathcal{G}^{\rm iso.}$).
For clarity, the cycle indices of the group generators, $\left\lbrace Z\right\rbrace$, are collected in the Table footnote\textsuperscript{a}.}
\resizebox{\linewidth}{!}{%
        \begin{tabular}{l l l   r r r    r r r   r r r   r r r  r r r} 
        \hline
$n$  & $I$ & $\mathcal{G}^{\rm iso.}$ & $X$ & C$_{X-2}$    & C$_{X-4}$  & C$_{X-6}$  & C$_{X-8}$  & C$_{X-10}$   & C$_{X-12}$ & C$_{X-14}$  & C$_{X-16}$ & C$_{X-18}$ & C$_{X-20}$ &C$_{X-22}$ &C$_{X-24}$& C$_{X-26}$&Total\textsuperscript{b}\\
     &     &      &     &  B$_1$N$_1$  &  B$_2$N$_2$&  B$_3$N$_3$&  B$_4$N$_4$&  B$_5$N$_5$  &  B$_6$N$_6$&  B$_7$N$_7$ & B$_8$N$_8$ & B$_9$N$_9$ & B$_{10}$N$_{10}$ & B$_{11}$N$_{11}$ & B$_{12}$N$_{12}$ &B$_{13}$N$_{13}$&\\
\hline
 2 & 1 & $D_2$   & 10 & 23 & 330    & 1,056   & 810       &         66 &             &      & & & & &&& 2,285\\
 3 & 2 & $D_2$   & 14 & 46 & 1,533  & 15,030  & 52,710    &     63,108 &      21,126 &         868 & & & & &&& 154,421 \\
 3 & 3 & $C_2$   & 14 & 91 & 3,024  & 30,030  & 105,210   &    126,126 &      42,112 &       1,716 & & & & &&& 308,309\\
 4 & 4 & $D_3$   & 18 & 51 & 3,096  & 61,890  & 510,888   &  1,837,836 &   2,859,726 &   1,750,320 &     328,500 & 8,110  & & &&& 7,360,417\\
 4 & 5 & $D_2$   & 18 & 77 & 4,644  & 92,848  & 766,332   &  2,756,964 &   4,289,544 &   2,625,760 &     492,750 & 12,190 & & &&& 11,041,109\\
 4 & 6 & $D_2$   & 16 & 63 & 2,785  & 40,142  & 225,690   &    504,894 &     421,050 &     103,140 &       3,290 &  & & &&& 1,301,054\\
 4 & 7 & $C_2$   & 18 & 153& 9,216  & 185,640 & 1,531,908 &  5,513,508 &   8,577,408 &   5,250,960 &     984,870 & 24,310 & & &&& 22,077,973 \\
 4 & 8 & $C_2$   & 18 & 154& 9,216  & 185,696 & 1,531,908 &  5,513,928 &   8,577,408 &   5,251,520 &     984,870 & 24,380 & & &&& 22,079,080 \\
 4 & 9 & $C_1$   & 18 & 306& 18,360 & 371,280 & 3,063,060 & 11,027,016 &  17,153,136 &  10,501,920 &   1,969,110 & 48,620 & & &&& 44,152,808\\
 5 & 10& $D_2$   & 22 & 116& 11,055 & 373,110 & 5,597,460 & 40,739,328 & 149,382,156 & 274,364,760 & 240,075,990 & 88,915,400 & 10,671,738 &  176,484 &&& 810,307,597 \\
 5 & 11& $D_2$   & 20 & 98 &  7,352 & 193,984 & 2,205,812 & 11,641,224 & 29,103,760 & 33,258,880 & 15,592,270 & 2,310,220 & 46,448 &  &&& 94,360,048 \\
 5 & 12--15,18& $C_2$   & 22 & 231 & 22,000 & 746,130 & 11,192,940 & 81,477,396 & 298,755,072 & 548,725,320 & 480,140,430 & 177,827,650 & 21,340,704 & 352,716  &&& 8,102,902,945\\
 5 & 16& $C_2$   & 22 & 232 & 22,000 & 746,220 & 11,192,940 & 81,478,656 & 298,755,072 & 548,729,520 & 480,140,430 & 177,830,800 & 21,340,704 & 352,968  &&& 1,620,589,542\\
 5 & 17& $C_2$   & 20 & 190 & 14,580 & 387,600 & 4,409,580 & 23,279,256 & 58,200,240  & 66,512,160  & 31,179,150  & 4,618,900   & 92,504 &  &&& 188,694,160 \\
 5 & 19,20,22-24& $C_1$   &22 & 462 & 43,890 & 1,492,260  &  22,383,900 & 162,954,792 & 597,500,904 & 1,097,450,640 & 960,269,310 &  355,655,300 & 42,678,636 &  705,432 &&& 16,205,677,630\\
 5 & 21& $C_1$   &20 & 380 & 29,070 & 775,200 & 8,817,900 & 46,558,512 & 116,396,280 & 133,024,320 & 62,355,150 & 9,237,800  & 184,756 & &&& 377,379,368\\
 6 & 25,26& $D_2$ & 26 & 163 & 22,542 & 1,151,216 & 27,343,030 & 334,640,790 & 2,230,954,440 & 8,286,315,840 & 17,090,574,930 & 18,989,469,950 & 10,634,147,524 & 2,636,560,416 & 219,721,684 & 2,600,612 & 120,907,006,274  \\
 6 & 27   & $D_2$ & 24 & 141 & 16,059 & 673,270 & 12,873,780 & 123,563,628 & 624,680,196 & 1,682,775,288 & 2,366,416,530 & 1,636,032,230 & 490,822,458 & 48,678,084 & 676,984 & & 6,987,208,648  \\
 6 & 28,29,34 \& 36& $C_2$   &24 & 276 & 31,944 & 1,345,960 & 25,742,970 & 247,118,256 & 1,249,329,312 & 3,365,515,296 & 4,732,773,210  & 3,272,028,760   & 981,616,944 & 97,349,616 & 1,352,540 && 55,896,820,336 \\
 6 & 30,31& $C_2$ &22 & 231 & 22,000 & 746,130 & 11,192,940 & 81,477,396 & 298,755,072 & 548,725,320  & 480,140,430  & 177,827,650  & 21,340,704  & 352,716    &&& 3,241,161,178\\
 6 & 32,33,35,37--41& $C_2$ & 26 & 325 & 44,928 & 2,302,300 & 54,681,770 & 669,278,610 & 4,461,874,560 & 16,572,613,200 & 34,181,059,770 & 37,978,905,250 & 21,268,222,976 & 5,273,104,200 & 439,431,356 & 5,200,300 & 967,253,756,360 \\
  6 & 42--45 & $C_2$ & 26 & 326 & 44,928 & 2,302,432 & 54,681,770 & 669,281,580 & 4,461,874,560 & 16,572,631,680 & 34,181,059,770 & 37,978,939,900 & 21,268,222,976 & 5,273,120,832 & 439,431,356 & 5,201,224 & 483,627,173,336 \\
 6 & 46--49,62,65,66 \& 76 & $C_1$ & 24 & 552 & 63,756 & 2,691,920 & 51,482,970 & 494,236,512 & 2,498,640,144 & 6,731,030,592 & 9,465,511,770 & 6,544,057,520 & 1,963,217,256 & 194,699,232 & 2,704,156 &  & 223,586,691,040  \\ 
 6 & 50--61,63,64,67--75 & $C_1$ & 26 & 650 & 89,700 & 4,604,600 & 109,359,250 & 1,338,557,220 & 8,923,714,800 & 33,145,226,400 & 68,362,029,450 & 75,957,810,500 &  42,536,373,880 & 10,546,208,400 & 878,850,700 & 10,400,600 & 5,561,704,201,450 \\ 
 7 & 77 & $D_6$  & 24 & 49 & 5,411 & 224,626 & 4,292,790 & 41,190,876 & 208,237,164 & 560,936,856 & 788,825,460 & 545,356,070 & 163,616,810 & 16,228,212 & 226,150 &  & 2,329,140,474 \\
\hline
\end{tabular}
}
\footnotesize \textsuperscript{a} $\left\lbrace Z\right\rbrace: \left\lbrace p_x^y,\,p_n^m p_l^k \right\rbrace$ 1) \{ $P_2^5, P_2^4$ \}, 2) \{ $P_2^7, P_2^6$ \}, 3) \{ $P_2^7$ \}, 4) \{ $P_3^6, 2P_2^9$ \}, 5) \{ $P_2^9, P_2^8$ \}, 6) \{ $P_2^8, P_2^6$ \}, 7) \{ $P_2^9$ \} , 8) \{ $P_2^8$ \}, 9) \{ $P_1^{18}$ \}, 10) \{ $P_2^{11}, P_2^{10}$ \}, 11) \{ $P_2^{10}, P_2^8$ \}, 12--15,18) \{ $P_2^{11}$ \}, 16) \{ $P_2^{10}$ \}, 17) \{ $P_2^{10}$ \}, 19,20,22--24) \{ $P_1^{22}$ \}, 21) \{ $P_1^{20}$ \}, 25,26) \{ $P_2^{13}, P_2^{12}$ \}, 27) \{ $P_2^{12}, P_2^{10}$ \}, 28,29,34) \{ $P_2^{12}$ \}, 30,31) \{ $P_2^{11}$ \}, 32,33,35,37--41) \{ $P_2^{13}$ \}, 36) \{ $P_2^{12}$ \}, 42--44) \{ $P_2^{12}$ \}, 45--49,62,65,66) \{ $P_1^{24}$ \}, 50--61,63,64,67--75) \{ $P_1^{26}$ \}, 76) \{ $P_6^4, P_2^{12} \}, P_2^{10}$ \};\,
\footnotesize \textsuperscript{b} For a given $n$, values are reported as sum over all $I$.
 \label{tab:allcarbons}
\end{table*}
\end{turnpage}

\section{\label{sec:level2}Results and Discussion}
\subsection{Clar structures of Benzenoid Hydrocarbons}
Benzenoid hydrocarbons (BHs) are a class of fusenes with two or more
six-membered rings that are mutually-condensed, {\it i.e.} 
each ring sharing an edge with at least another ring\cite{balaban1968chemical}. 
The structures of BHs are typically isomorphic to that of hexagonal sub-lattices, however, helicenoid compounds---a class of BHs---adopt structures that are non-isomorphic to hexagonal sub-lattices\cite{brinkmann2007fusenes}. 
The valence electronic properties of BHs can be explained by Clar aromatic sextet theory\cite{clar1964c,sola2013forty}. 
The essence of this theory is to derive a graphical representation of the electronic structure, which  encodes bonding information from all the Kekul{\'e} structures (KSs) as well as the resonant hybrid one. For a given BH every KS can be represented as a Clar diagram composed of one or more aromatic $\pi$-sextets (hexagons inscribed with circles) and non-sextet carbon-carbon double bonds. A Clar diagram with the most number of aromatic $\pi$-sextets is the so-called {\it Clar structure} (CS). It is important to consider the CS as a versatile chemical descriptor for benzenoid-Kekulean structures. Another such descriptor is Wiener index\cite{nikolic1995wiener,vukivcevic2004wiener}, which is a topological score  mapping the molecular structure to various chemical properties.
Algebraic representations of CSs have indeed been shown to capture more chemical information than a single KS\cite{randic2004algebraic,randic2018local}. 
Multiple Clar diagrams, where the $\pi$-sextets are located in adjacent rings, can be collectively represented by a single {\it migrating} CS\cite{sola2013forty}. 
The total number of $\pi$-sextets in the CS has been shown to correlate with bond-length variations, aromaticity and also HOMO-LUMO gap of the BH\cite{sola2013forty}. 
Throughout this paper, all molecular cartoons are presented in the Clar form.

\subsection{Constructive Enumeration of Benzenoid  Hydrocarbons}
KSs of PAHs with up to six benzene rings have been constructively enumerated using an in-house program. For a given number of rings, the program exhaustively generates all possible molecular structures by tiling ({\it i.e.} plane-filling) and ignores all the disconnected structures such as non-bonded molecular dimers. As an exemplary case, in Fig.~\ref{fig01} we have collected all the generated structures containing three rings. The resulting structures include benzenoid ($B$) as well as non-benzenoid ($nB$) compounds. While the latter are excluded in this study, among benzenoids we have considered only those with a valid Kekul{\'e} formula ($K$). We note in passing that compounds which do not have a valid Kekul{\'e} formula ($nK$) such as the one shown in Fig.~\ref{fig01} are non-aromatic as long as the system is charge neutral. It is worthwhile to note that the aforementioned procedure allows certain benzenoid structures with more rings to qualify, for instance, isochrysene a formally 4-ring system (see Fig.~\ref{fig01}, blue labels) can be generated using three benzene rings mutually connected at 1,2 sites in a non-benzenoid fashion ({\it i.e.} with sigma bondings). 
For a given number of rings, we  exclusively collect all the benzenoid molecules with valid Kekul{\'e} formulae ($B+K$) (see Fig.~\ref{fig01}, green labels). 
The structures of all the resulting hydrocarbons comprising up to six benzene rings are displayed in Fig.~\ref{PAH}. For 2-, 3-, 4-, 5-, and 6-ring compounds, our procedure has resulted in 1, 2, 6, 15, and 52 hydrocarbons, respectively. These numbers agree with those given by Brinkmann {\it et al.}\cite{brinkmann2007fusenes} who further enumerated the hydrocarbons with more benzene rings and reported 195, 807, 3513, 16025 entries for 7-, 8-, 9-, and 10-ring compounds, respectively. We note that our plane-filling algorithm does not distinguish between the 6-ring compound hexahelicene (structure 45 in Fig.~\ref{PAH}) and the peri-condensed 7-ring compound coronene a.k.a. superbenzene (structure 77 in Fig.~\ref{PAH})---both 3D structures surjectively mapped to the same plane-filling 2D pattern. For this reason, we have included coronene as the sole 7-ring system in this work. It is important to note that the helical compounds such as tetrahelicene (8 in Fig.~\ref{PAH}), pentahelicene (13 in Fig.~\ref{PAH}) and hexahelicene (45 in Fig.~\ref{PAH}) are chiral with two enantiomers having same substitution pattern. For a given number of rings we found the number of hydrocarbon compounds in Fig.~\ref{PAH} to fully agree with a constructive enumeration performed using the program {\tt CaGe} \cite{brinkmann2010cage}. The main objective of this study concerns the grouping of seventy-seven PAHs shown in Fig.~\ref{PAH} according to their molecular symmetry group as well as available substitution sites and enumerate the number of all possible unique compounds that can be formed by replacing pairs of C atoms in PAHs with B and N atoms.

\subsection{\label{sec:level3}Chemical Space Enumeration: Mathematical Formulation and Computation}
Combinatorial enumeration of isomers, topologies, nuclear spin statistical weights, etc., can be efficiently performed using generalized character cycle index (GCCI)\cite{balasubramanian1985applications,balasubramanian1992combinatorics} of the molecular symmetry group that are subgroups of the complete nuclear permutation inversion (CNPI)\cite{bunker2006molecular} group. 
The GCCI ($\mathcal{Z}^{\Gamma}_\mathcal{G}$) of a group $\mathcal{G}$ corresponding to a particular irreducible representation $\Gamma$ is defined as
\begin{eqnarray}
\mathcal{Z}^{\Gamma}_\mathcal{G} & = & \frac{1}{ |\mathcal{G}| }
\sum_{g \in \mathcal{G}} \chi^{\Gamma}(g)
\left[ \Pi_{k} \mathcal{F}_k \right]
\label{eq:gcci}
\end{eqnarray}
where $g$ goes over all elements of the permutation group $\mathcal{G}$, $\chi^{\Gamma}(g)$ is the character of $g$'s matrix representation for the given irreducible representation $\Gamma$ and 
$\Pi_{k} \mathcal{F}_k$ corresponds to the cycle representation of element $g$ with $k$ running over all the factor-cycles of $g$. 
In Eq.~\ref{eq:gcci}, 
$|\mathcal{G}|$ denotes the group order and
the figure counting series $(\mathcal{F})$ for cycle length $n$ is given by $\mathcal{F} = \mathcal{A}^n + \mathcal{B}^n + \ldots$ with $\mathcal{A}$, $\mathcal{B}$, etc. being the objects to be permuted. For the special case where $\Gamma$ is the (fully) symmetric representation of $\mathcal{G}$, one arrives at the P{\'o}lya enumeration formula\cite{polya1937kombinatorische,polya2012combinatorial} that has been so successfully employed for enumerating various chemical subspaces\cite{faulon2005enumerating,balaban1991enumeration}.
\begin{eqnarray}
\mathcal{Z}_\mathcal{G} & = & \frac{1}{ |\mathcal{G}| }
\sum_{g \in \mathcal{G}} 
\left[ \Pi_{k} \mathcal{F}_k \right]
\label{eq:ci}
\end{eqnarray}

For the enumeration of BN-PAH chemical space, $\mathcal{F}$ takes the form $C^n+B^n+N^n$, where $n$ is the cycle length. Application of Eq.~\ref{eq:ci}, for instance, to naphthalene (${\rm C}_{10}{\rm H}_{8}$) with $\mathcal{G}$ isomorphic to the ${D}_2$ point group results in the following pattern inventory
\begin{widetext}
\begin{eqnarray}
\mathcal{Z}_\mathcal{G} & = & C_{10}+3C_9N+3C_9B+15C_8N_2+23C_8NB+15C_8B_2+32C_7N_3+92C_7N_2B+ \nonumber  \\ 
                        & & 92C_7NB_2+32C_7B_3+60C_6N_4+212C_6N_3B+330C_6N_2B_2+212C_6NB_3+ \nonumber  \\ 
                        & & 60C_6B_4+66C_5N_5+318C_5N_4B+636C_5N_3B_2+636C_5N_2B_3+318C_5NB_4+ \nonumber \\ 
                        & & 66C_5B_5+60C_4N_6+318C_4N_5B+810C_4N_4B_2+1056C_4N_3B_3+810C_4N_2B_4+\nonumber \\
                        & & 318C_4NB_5+60C_4B_6+32C_3N_7+212C_3N_6B+636C_3N_5B_2+1056C_3N_4B_3+\nonumber \\
                        & & 1056C_3N_3B_4+636C_3N_2B_5+212C_3NB_6+32C_3B_7+15C_2N_8+92C_2N_7B+\nonumber \\
                        & & 330C_2N_6B_2+636C_2N_5B_3+810C_2N_4B_4+636C_2N_3B_5+330C_2N_2B_6+\nonumber \\
                        & & 92C_2NB_7+15C_2B_8+3CN_9+23CN_8B+92CN_7B_2+212CN_6B_3+318CN_5B_4+\nonumber \\
                        & & 318CN_4B_5+212CN_3B_6+92CN_2B_7+23CNB_8+3CB_9+N_{10}+3N_9B+15N_8B_2+\nonumber \\
                        & &32N_7B_3+60N_6B_4+66N_5B_5+60N_4B_6+32N_3B_7+15N_2B_8+3NB_9+B_{10}
\label{eq:patterninventory}
\end{eqnarray}
\end{widetext}
where each term corresponds to the stoichiometry ${\rm C}_{x}{\rm B}_{y}{\rm N}_{z}$ ($x,y,z\in{0,\ldots,10}$ and $x+y+z=10$). 
In all cases, the number of ${\rm H}$ atoms, not shown in Eq.~\ref{eq:patterninventory},
is 8 as in the parent hydrocarbon naphthalene. 
The coefficient of each term in Eq.~\ref{eq:patterninventory} gives the number of unique compounds for that stoichiometry ({\it i.e.} the number of constitutional isomers). For example, the total number of naphthalene derivatives with stoichiometry ${\rm C}_{8}{\rm B}{\rm N}{\rm H}_{8}$ is 23 as given by the fifth term on the right side of  Eq.~\ref{eq:patterninventory}.
For asymmetric molecules, the permutation group is isomorphic to the $C_1$ point group and the GCCI reduces to the multinomial expansion 
\begin{equation}
\mathcal{Z}_{C_{1}}= \left( C+B+N \right)^m = \underset{x+y+z=m}{\sum}\left(\begin{array}{c}
m\\
x,y,z
\end{array}\right)C_{x}B_{y}N_{z},
\label{eq:c1}
\end{equation}
where $m$ is the number of sites available for substitution and the summation goes over all combinations of non-negative $x$, $y$ and $z$ with the constraint $x+y+z=m$.

In the present work, we are solely interested in the enumeration of isosteric/isoelectronic compounds with equal number of B and N atoms. For example, all naphthalene based compounds are of the form
${\rm C}_{10-2y}{\rm B}_{y}{\rm N}_{y}{\rm H}_{8}$, where $y=0,\ldots,5$. For all seventy-seven hydrocarbons listed in Fig.~\ref{PAH}, we have separately collected the pattern inventory for such substitutions of the inner C atoms (Table~\ref{tab:innercarbons}), outer C atoms (Table~\ref{tab:PeripheralCarbons}), and both inner and outer C sites without restrictions (Table~\ref{tab:allcarbons}). 
In these tables, we have also listed the cycle indices of the group generators denoted by $\left\lbrace Z\right\rbrace$. 
For naphthalene, in the case of substitution of the two inner C sites, the relevant group is $\mathcal{G}=\left\lbrace (1)(2), (1,2) \right\rbrace$ and its generator is $\left\lbrace (1,2) \right\rbrace$. The cycle index set is then denoted by $\lbrace P_2^1 \rbrace$, {\it i.e.}, one cycle of length two. 
For the substitution pattern of the peripheral C atoms of naphthalene, the appropriate permutation group $\mathcal{G}$ can be formed using the generator set $\left\lbrace (1,9)(2,8)(3,7)(4,6), (1,4)(2,3)(6,9)(7,8) \right\rbrace$ with $\left\lbrace Z\right\rbrace=\lbrace 2P_2^4 \rbrace$ (see Table~\ref{tab:PeripheralCarbons}). 
The permutation group used for collecting the substitution pattern of all ten carbons of naphthalene is derived using the generator set $\left\lbrace (1,9)(2,8)(3,7)(4,6), (1,4)(2,3)(5,10)(6,9)(7,8) \right\rbrace$ with the corresponding cycle indices $\lbrace P_2^5, P_2^4 \rbrace$, {\it i.e.}, a generator with five indices of length two, and another with four indices of length two. 
For all seventy-seven PAHs displayed in Fig.~\ref{PAH}, the total number of compounds obtained by substituting all (both inner and outer C sites without restrictions) available C atoms is graphically presented in Fig.~\ref{fig:circular}. The overall trend is that for a given number of substitution sites maximal number of products is noted for asymmetric hydrocarbon frameworks that follow a multinomial distribution, see Eq.~\ref{eq:c1}. 






\subsection{Symmetry-controlled yield-pattern selectivity}

\begin{figure}[htbp!]
\centering
\includegraphics[width=8.5cm]{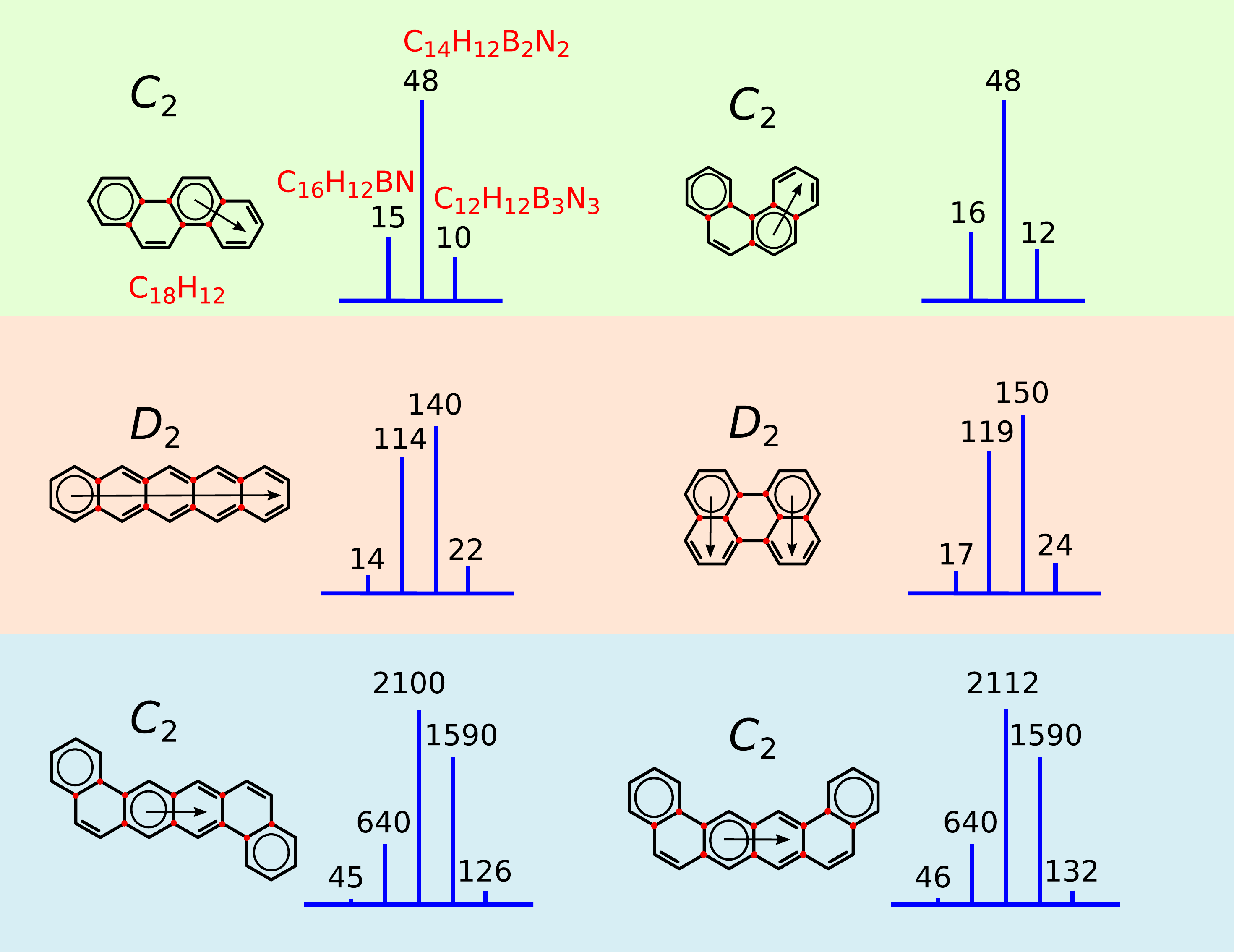} 
\caption{
Selected cases of symmetry-controlled yield-pattern selectivity across
BN-PAH compounds. Molecules on left and right sides have the same point group symmetry and same number of inner C sites available for substitution.
}
\label{fig:isomerdist}
\end{figure}

For a given PAH, deviations from a multinomial distribution of B, N-substituted compounds increase with the order of the symmetry group of the hydrocarbon. Such a trend has been discussed in the past for the enumeration of gallium arsenide (Ga$_x$As$_y$) clusters\cite{balasubramanian1988enumeration,balasubramanian1992combinatorics}. In the case of symmetric PAHs, it is interesting to note that even when two (or more)
PAHs share the same rotational group and comprise of same number of substitution sites ($X$), 
these compounds can lead to distinct GCCIs ($\mathcal{Z}_\mathcal{G}$), therefore distinct product distributions. For example, see the two rows corresponding to $n=4,\,I=7$ and $n=4,\,I=8$ in Table~\ref{tab:innercarbons}. 
While on one hand, the symmetry groups of the respective compounds, chrysene and tetrahelicene (7 and 8 in Fig.~\ref{PAH}) correspond to the same isomorphic point group $C_2$; on the other, their inner-substituted analogues result in different constitutional isomer distributions (see Table~\ref{tab:innercarbons}). Similar trends are noted also in the cases of larger PAHs; few illustrative examples are on display in  Fig.~\ref{fig:isomerdist} while all such instances are collected in Table~\ref{tab:dist_table}. 
Qualitatively, such a selectivity can be understood by considering the fact that, in the case of tetrahelicene (a {\it cisoid} fused system, structure 8 in Fig.~\ref{PAH}), two C atoms lie on the principal axis whose positions are invariant with respect to the $C_2$ rotation. In other words, the corresponding labels form an {\it invariant subspace} of the isomorphic permutation operation. 
In the case of chrysene (a {\it transoid} fused system, structure 7 in Fig.~\ref{PAH}), the $C_2$ principal axis is normal to the molecular plane and 
passes through the geometric center of the molecule without coinciding with any of the C atoms, {\it i.e.}, zero-invariant subspace. 
So tetrahelicene, which has a larger invariant subspace, can be thought of as a permutationally less symmetric molecule than chrysene, as far as the inner C atom framework is considered.


\begin{table}[htbp!]
    \centering
     \caption{
     Symmetry-controlled yield-pattern selectivities
     across all seventy-seven PAHs: 
     $n$ is the number of rings in the parent hydrocarbon,
      $\mathcal{G}$ is the point group isomorphic to the symmetry group, 
      $X$ is the number of substitution sites in the parent hydrocarbon, 
$I$ is the index of the hydrocarbon in Fig.~\ref{PAH},
and $\lbrace Z \rbrace$  denotes the cycle indices of the generator subgroup.}
    \begin{tabular}{l l l  l l  l l  l}
    \hline
    $n$~~~& $\mathcal{G}$~~~  & $X$~~~ & \multicolumn{2}{l}{$\mathcal{A}$} &~~~& \multicolumn{2}{l}{$\mathcal{B}$}\\ 
    \cline{4-5}\cline{7-8}
        & & & $I$ &$\lbrace Z \rbrace$& & $I$ &$\lbrace Z \rbrace$ \\
    \hline
    \multicolumn{8}{l}{Inner sites} \\
    4 & $C_2$ & 6 & 7 &\{$P_2^3$\} && 8 & \{$P_2^2$\} \\
    5 & $D_2$ & 8 & 10 &\{2$P_2^4$\} && 11 & \{$P_2^4,P_2^2$\} \\
    6 & $C_2$ & 10 & 28--41 &\{$P_2^5$\} && 42--45 & \{$P_2^4$\} \\
        \multicolumn{8}{l}{} \\

    \multicolumn{8}{l}{Peripheral sites} \\
    5 & $C_2$ & 14 & 12--15, 18 &\{$P_2^7$\} && 16 & \{$P_2^6$\} \\
            \multicolumn{8}{l}{} \\

    \multicolumn{8}{l}{All sites} \\
    4 & $C_2$ & 18 & 7 &\{$P_2^9$\} & &8 &\{$P_2^8$\} \\
    5 & $C_2$ & 22 & 12--15, 18 &\{$P_2^{11}$\} & &16 & \{$P_2^{10}$\} \\
    6 & $C_2$ & 26 & 32, 33, 35, 37--41 &\{$P_2^{13}$\} & & 42--45 &\{$P_2^{12}$\} \\
    \hline
    \end{tabular}
     
    \label{tab:dist_table}
\end{table}
To gain a more quantitative appreciation of the 
aforementioned selectivity and more importantly, to compute the actual
distribution, one has to consider the corresponding nuclear permutation groups and inspect the cycle length structure of the group generators. Let us consider the GCCIs for the case
of inner-site substitutions of the two PAHs chrysene 
($\mathcal{Z}_\mathcal{G,A}$) and tetrahelicene ($\mathcal{Z}_\mathcal{G,B}$):
\begin{eqnarray}
\mathcal{Z}_\mathcal{G,A} & = & \frac{1}{2} \lbrace (B+N+C)^6 + (B^2+N^2+C^2)^3 \rbrace \\
\mathcal{Z}_\mathcal{G,B} & = & \frac{1}{2} \lbrace
(B+N+C)^6 + (B+N+C)^2(B^2+N^2+C^2)^2 \nonumber \rbrace \\ 
& & 
\label{eq.symm}
\end{eqnarray}
While expanding the expressions on the right side yields the pattern inventory, it is evident that the difference in the number
of substituted products between the two PAHs arises from the generator cycle length structure. In Eq.~\ref{eq.symm}, the first factor on the second term, $(B+N+C)^2$, indicates that two C atoms are invariant with respect to the corresponding symmetry element $C_2\equiv(1)(2)(3,4)(5,6)$. Typically, BN-PAH molecules are synthesized by starting with suitable precursor compounds, see for example \Ref{bosdet200710a}. 
In order to realize the symmetry-controlled yield-pattern selectivity proposed here, for any given PAH, synthesis strategies must statistically account for all possible substituted compounds.

\subsection{High-throughput first-principles modeling of BN-PAH compounds}
All the 7,453,041,547,842, {\it i.e.} $7.4$ tera BN-PAH molecules enumerated in the previous
sections feature even number of electrons and are of closed-shell type. 
Past computational investigations of some of these compounds have demonstrated the reliability of density functional approximations (DFAs) 
for semi-quantitative prediction of their structures and 
electronic properties\cite{marcon2007tuning,al2014water,ghosh2011density}. It is important to note that while first-principles modeling of any single constituent of the BN-PAH dataset presents no conceptual challenge, the sheer size of this dataset will render any brute-force high-throughput computational endeavor aiming towards complete coverage impractical---even when depending on petascale computer facilities. 
For instance, geometry relaxation of a typical medium-sized organic molecule using even a semi-empirical method such as PM7 requires about 10 CPU seconds. Carrying out such calculations for all the  $7.4$ tera BN-PAH molecules would require over two million CPU years. Deploying thousands of CPU cores for this purpose can at most
decrease this time by three orders of magnitude.

 In this first study, to gain rational insights into the stability and optoelectronic properties of BN-PAH compounds, we have performed high-throughput DFT modeling for only a representative subset. To this end, we restrict the number of molecules to a feasible size by considering all possible substitutions in naphthalene (row 1 of Table \ref{tab:allcarbons}) and only single B,~N pair substitutions in the remaining 76 PAHs (column 5 of Table \ref{tab:allcarbons}), overall amounting to 33,059 (33k) compounds. 
For all these compounds we have performed geometry optimizations and collected various properties for minimum energy structures; details of the DFT calculations are provided in Computational Methods.
In the following, we report on the geometric features of the BN-PAH molecules and discuss deviations of the predicted structures from those expected solely based on formal hybridization scenario of the unsubstituted PAH compounds. Then we present the distribution of HOMO-LUMO gap of the 33k compounds in the context of solar spectrum. Finally we comment on inter-property correlations between dipole moment, electronic gap and atomization energy
which are relevant to rational compound design strategies.
 
\subsubsection*{E.1. Deviations from ideal $sp^2$ geometry}
\begin{figure}[htb!]
\includegraphics[width=8.5cm]{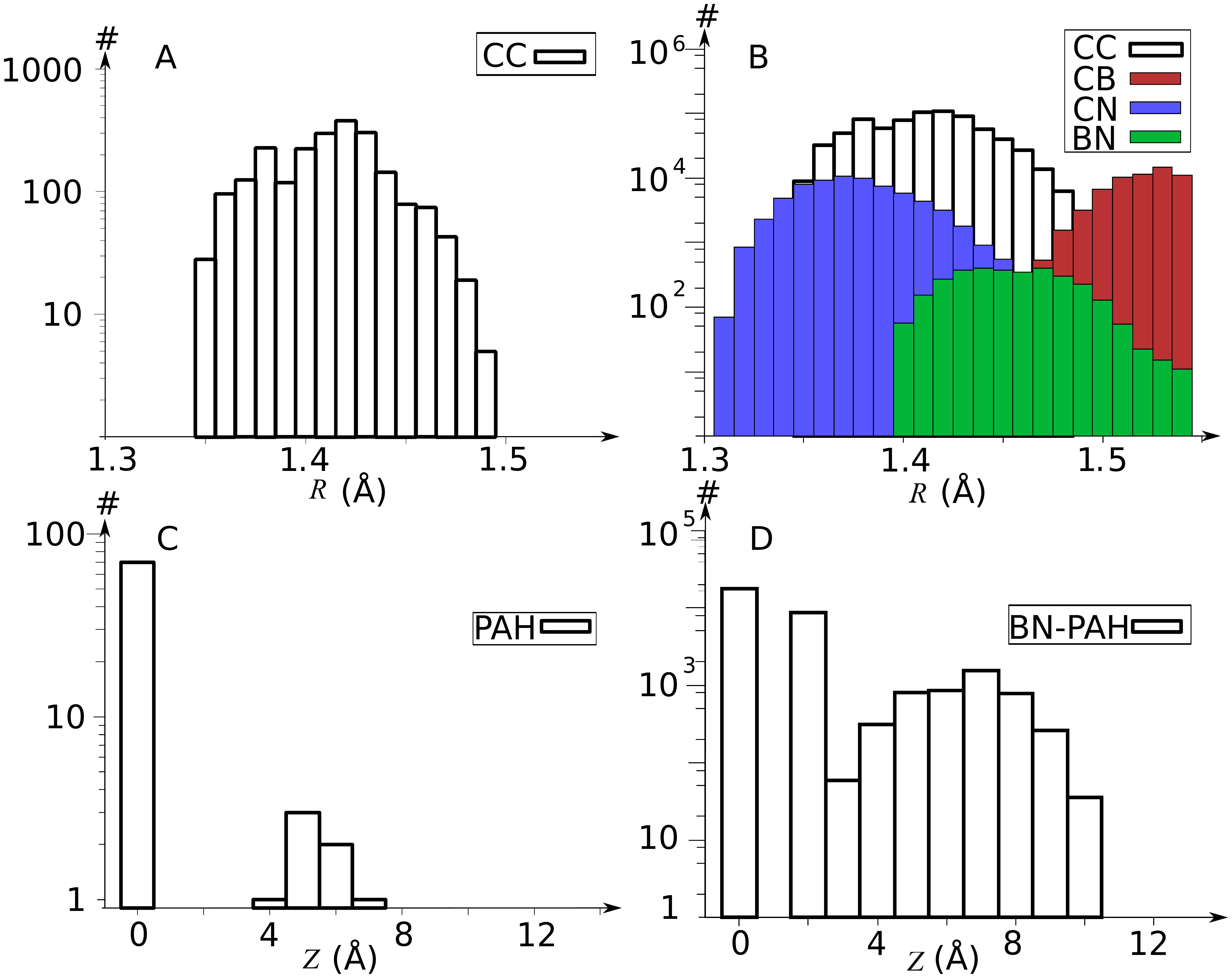}
\caption{ Comparison of trends in bond lengths ($R$) and out-of-plane deviation measures ($Z$) between seventy-seven PAHs and their B, N-substituted counterparts. Bond length distributions are shown in panels A and B, while that of the out-of-plane deviations ($Z$) are shown in panels C and D. Out of 33,059 BN-PAH compounds studied using DFT, only 30,797 compounds with a single B,N pair have been considered
for a comparison across PAHs, the remaining 2,262 molecules are naphthalene derivatives with multiple B,~N pairs.  See text for more details.}
\label{DFTdistoop}
\end{figure}

Formally, we understand the aromatic character of benzenoid 
species by Clar's  $\pi$-sextet rule \cite{clar1972aromatic,feixas2008performance,sola2013forty},
which predicts hydrocarbons with migrating sextets to contain bonds of equivalent lengths and a planar geometry, characteristic of an ideal $sp^2$ hybridization. On the other hand, compounds with fewer $\pi$-sextets show local aromaticity resulting in deviations from  ideal $sp^2$ structures. 
Further, trends in thermodynamic stability are also expected to correlate with the number of aromatic $\pi$-sextets---the larger its value, greater should be the stability. However, it is important to note that all these assumptions that are expected to hold for PAHs need not hold for the hetero-compounds. To gain a first-principles understanding of key geometric features of both the PAHs as well as their hetero analogues, we have collected bond length distributions and out-of-plane deviations (OD) from DFT calculations in Fig.~\ref{DFTdistoop}. 
The OD was obtained by finding a best-fit molecular plane through rigid body rotations, and then calculating the deviations along the normal ($z$-axis) in a root-mean-squares fashion,  $Z=\sqrt{\sum_{i=1}^N (\Delta z)^2}$, where $N$ is the number of atoms.
In order to compare these structural features across various PAHs, we consider only those substituted compounds containing only one pair of B,~N atoms resulting in 30,797 (31k) BN-PAH compounds. 

First of all, let us inspect the CC bond distances of the hydrocarbons in Fig.~\ref{DFTdistoop}A and note the bond lengths to vary between the values of a conventional CC double bond ($\sim$ 1.34~\AA) and a CC single bond ($\sim$ 1.5~\AA). In the case of BN-PAH compounds (see Fig.~\ref{DFTdistoop}B), we note the CN double bond to display the widest distribution followed by BN and CB double bonds. 
This variation of bond lengths implies the  heteroatoms to introduce structural inhomogeniety thereby weakening electron delocalization across the molecule. Such an effect may
be understood via the electronegativity criterion alone, where one can classify the CN moiety to be electron-rich and CB moiety to be electron-deficient compared to the CC fragment resulting in a net gain in electron density on the CN fragments.
Moreover, even though the BN fragment is isoelectronic and isosteric to the CC one, due to the larger ionic character, we find the average BN bond length to be larger than the average CC bond length (Fig.~\ref{DFTdistoop}B). 




\begin{figure}[htbp!]
\centering
\includegraphics[width=8.0cm]{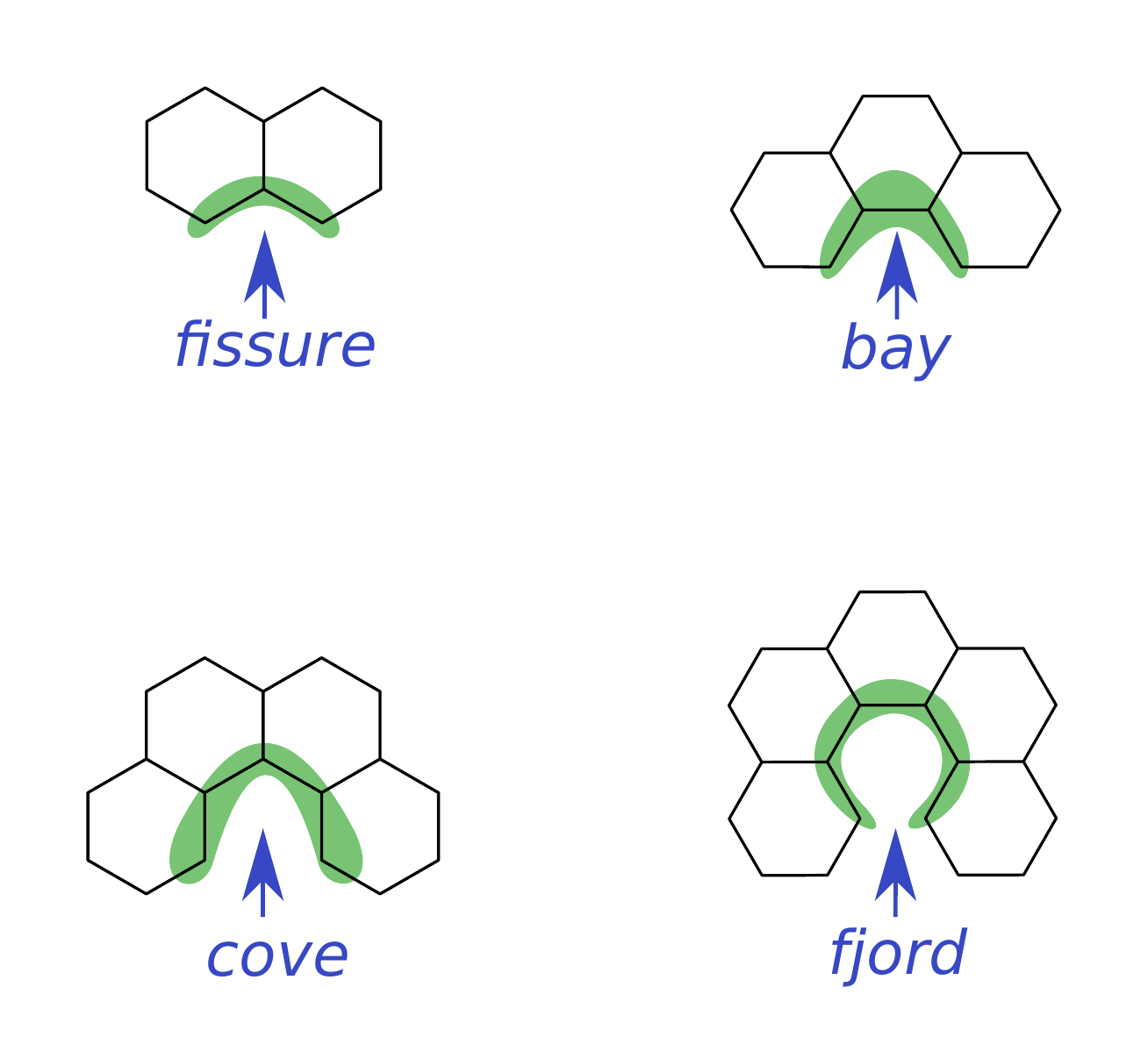} 
\caption{Various topological features as encountered in 
the smallest polycyclic aromatic hydrocarbons exhibiting them. For clarity only one occurrence of {\it fissure} is shown.}
\label{topfeat1}
\end{figure}

Moving on to the OD of the hydrocarbons and their hetero counterparts, it is
useful to note that such distortions signify the presence of strain. For a better clarity, it is worthwhile to recall the formal classification of benzenoid compounds based on the presence of the various topological features--{\it fissure}, {\it bay}, {\it cove} and {\it fjord}\cite{dias2005perimeter,pogodin2002overcrowding}; see Fig.~\ref{topfeat1}. The bay, cove and fjord regions have H atoms in close proximity introducing strain in a strictly planar geometry. A cove may be thought to be comprised of two proximate bays while a fjord three proximate bays. For a given number of C atoms, a strictly peri-condensed compound---approaching a more circular topology---must have minimum number of external C atoms, hence fewer number of bays\cite{dias1990periodic}. Evidently, structures with a fjord region are more susceptible to OD followed by ones with coves and lastly by those with bays. 

The most striking feature in Fig.~\ref{DFTdistoop}C 
is that 90\% of the parent PAHs (70 out of 77) are perfectly planar, while a few do show moderate OD; such structures are those with fjords (13, 35, 45, 61, 64 \& 67 in Fig. \ref{PAH}) and multiple coves (26 in Fig. \ref{PAH}). 
In contrast, in Fig.~\ref{DFTdistoop}D, we note only about 66\% of the hetero structures (about 20k out of 31k) to be planar. To further quantify, let us consider a threshold of $Z=0.01$ \AA. 
While only 9.1\% of the PAHs have a larger $Z$ with respect to this threshold, 19.3\% of the B,~N-substituted ones have $Z>0.01$ \AA.

These results show that B,~N-substitution induces the molecules to distort from planar configurations. 
Loss of planarity also compromises the efficiency of these molecules as chromophores, or components of singlet-fission systems. Singlet-fission is a process by which an organic chromophore in an excited singlet state transfers its excess energy to a neighbouring chromophore in the ground state resulting in two triplet states
there by doubling the number of carrier charges\cite{smith2010singlet}. In inter-molecular singlet-fission, especially in molecular solids, planarity of the constituent molecules is crucial to stabilize the crystal via $\pi-$stacking\cite{bhattacharyya2017polymorphism}. Out of the 31k BN-PAH compounds 80.7\% satisfy this prerequisite. However, in actual 
singlet-fission applications, the singlet-triplet energetics also play a vital role. Therefore, more efforts are needed for a better evaluation of the structure-energetics trade-off across the BN-PAH dataset.

\subsubsection*{E.2. Trends in electronic structure}
\begin{figure}[htb!]
\includegraphics[width=8cm]{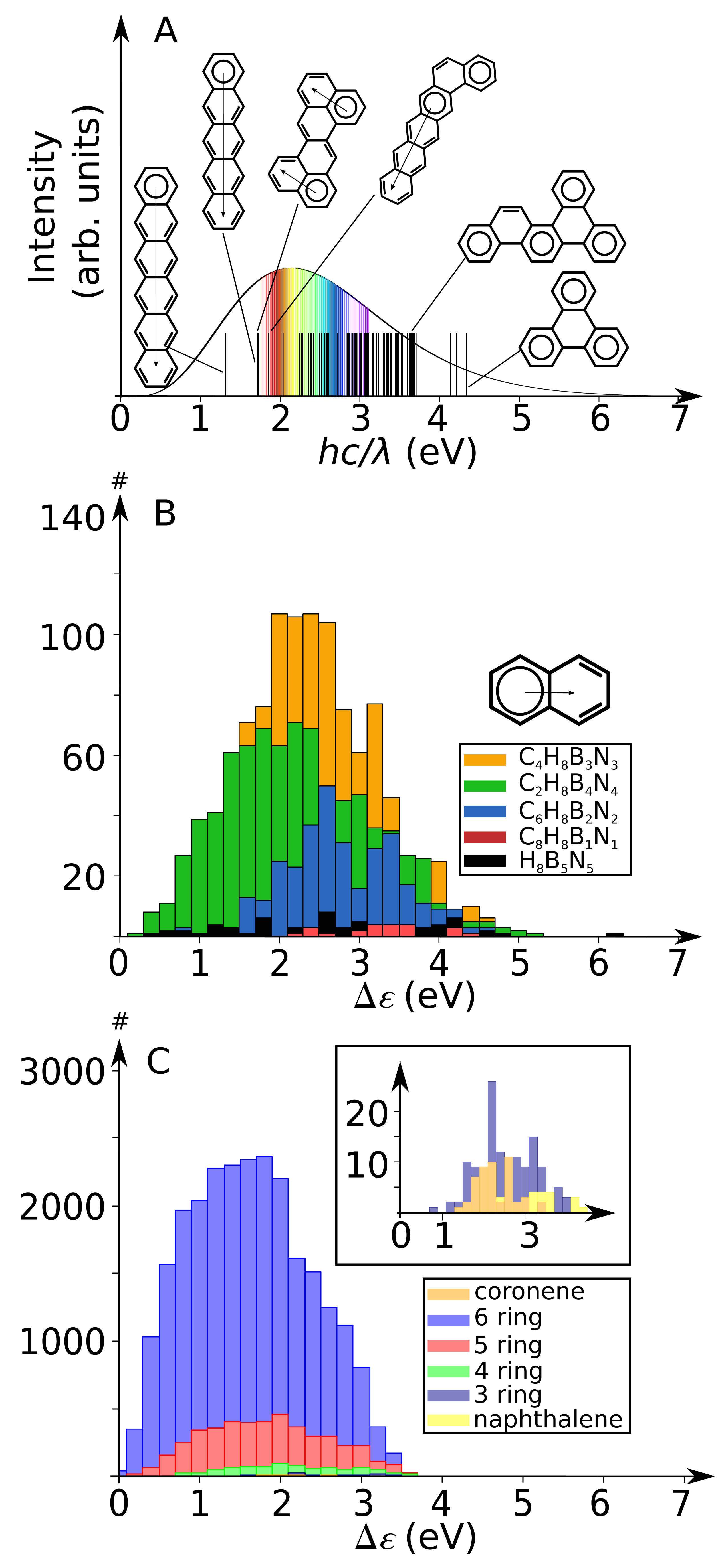}
\caption{Spectral distribution of HOMO--LUMO gap ($\Delta\varepsilon$) of 33,059 BN-PAH compounds. A) $\Delta\varepsilon$ of unsubstituted PAHs (77) mapped with selected Clar structures.
B) $\Delta\varepsilon$ of 2,285 (B,N)$_x$-substituted ($x=1,\ldots5$) isomers of naphthalene. C) $\Delta\varepsilon$ of 30,797 (B,N)$_1$-substituted isomers of all  PAHs; the inset zooms into under-represented compounds. The ideal solar spectrum with colors in the visible region is shown in panel-A for comparison.
}
\label{DFTgap}
\end{figure}

Clar structures also provide 
information about the electronic energy level separations of BHs. In general, with increasing number of $\pi$-sextets in the Clar structure, the HOMO-LUMO transition is blue-shifted\cite{sola2013forty}. It is interesting to note that this
formal empirical rule is corroborated by TPSSh results as seen in Fig.~\ref{DFTgap}A.
Out of seventy-seven hydrocarbons, most have HOMO--LUMO gaps ($\Delta\varepsilon$)
in the visible region of the solar spectrum (see Fig.~\ref{DFTgap}A). 
At the longer wavelength region near the spectral maximum, only about half-a-dozen elongated hydrocarbons are active. 
For the linear PAHs---napthalene, anthracene, tetracene, pentacene and hexacene---the TPSSh values deviate from the experimental counterparts\cite{george1968intensity,malloci2007time,malloci2011electronic} 
({\it i.e.} experimental$-$TPSSh) by  0.24, -0.69, 0.47, 0.60, and 0.57 eV, respectively amounting to an average prediction error of 0.24 eV and a root mean square error of 0.54 eV. These error measures imply the above discussed trends in HOMO--LUMO gaps to retain their semi-quantitative accuracy at least for other PAH molecules.
It may be noted that a number of solar cell applications have been based on the organic dye coumarin because of its very desirable electronic gap at 2.25 eV (552 nm) \cite{mishra2009metal}. However, none of the PAHs exhibit $\Delta\varepsilon$ in the yellow region of the spectrum, where the spectral energy density of the solar black-body radiation is maximum. 
It is important to note that for these transitions to be allowed, the corresponding transition dipole moment integrals must also be non-vanishing. Computation of such integrals must be done using time-dependent DFT; such efforts are beyond the scope of the present study. 
\begin{table}[!htbp]
    \centering
        \caption{Percentage distribution of HOMO--LUMO gap ($\Delta\varepsilon$) of 33,059 BN-PAH compounds across UV, visible and IR regions of the solar spectrum.}
    \begin{tabular}{l c c c }
    \hline
          Compounds    &               & \%($\Delta\varepsilon$) \\
              \cline{2-4}
   & IR                & ~~~~~~~~~~~~visible~~~~~~~~~~~~ & UV  \\
   &  $<$ 1.77 eV      &   1.77--3.09 eV       &  $>$ 3.09 eV \\
        \hline
            \multicolumn{4}{c}{naphthalene} \\
            C$_8$B$_1$N$_1$H$_8$ &  0.00 & 30.43 & 69.57 \\
            C$_6$B$_2$N$_2$H$_8$ &  4.85 & 56.97 & 38.18 \\
            C$_4$B$_3$N$_3$H$_8$ & 16.95 & 59.19 & 23.86 \\
            C$_2$B$_4$N$_4$H$_8$ & 29.14 & 49.01 & 21.85 \\
                 B$_5$N$_5$H$_8$ & 19.70 & 37.88 & 42.42 \\
            \multicolumn{4}{c}{larger rings} \\
             3 rings  &  7.30 & 62.04 & 30.66 \\
            4 rings  & 28.86 & 55.60 & 15.55 \\
            5 rings  & 43.14 & 49.36 &  7.50 \\
            6 rings  & 53.59 & 42.88 &  3.53 \\
            coronene  &  4.08 & 87.76 &  8.16 \\
        \hline
        \end{tabular}
    \label{tab:dist_table1}
\end{table}

Fig.~\ref{DFTgap}B presents the $\Delta\varepsilon$ of 2,285 (B, N)$_x$-substituted isomers of naphthalene and Fig.~\ref{DFTgap}C shows the $\Delta\varepsilon$ of 30,797 (B, N)$_1$-substituted isomers for all PAHs. In both figures, 
we observe the property distributions to follow roughly a Gaussian-type trend. Such statistical trends often imply that the character of the electronic excitation is preserved across all the constitutional isomers with same stoichiometry\cite{ramakrishnan2015electronic}.  

In Table \ref{tab:dist_table1} we have collected the gap distribution across different regions of solar spectra for (B,~N)$_1$-substituted compounds and all possible isomers of naphthalene. We observe that for most of the classes, majority of the molecules ($>50\%$) lie in the visible region of the solar spectrum. 
An interesting correlation can be drawn based on works by Hoffmann \textit{et al.} \cite{hoffmann1964extended,alkaabi2012ionic,niedenzu2012boron,zeng2014seeking}, where the authors explain the properties of a substituted PAH with B$_x$N$_y$ units 
as the consequence of perturbation of parent hydrocarbon's properties by heteroatoms.
In Table \ref{tab:dist_table1} we note that for naphthalene, the (B,~N)$_1$ substitution shifts from predominantly UV to predominantly UV-visible with increase in the number of heteroatoms, a maximum shift to the visible region is noted for the stoichiometry C$_4$B$_3$N$_3$H$_8$.


When comparing the modulation of $\Delta\varepsilon$ by (B,~N)$_1$ substitution (Fig.~\ref{DFTgap}C)
in all the PAHs with those of (B,~N)$_x$ substitution in naphthalene (Fig.~\ref{DFTgap}B), one notes a similar of spread of 0--4 eV in both cases with too few examples with $\Delta\varepsilon>4$ eV. 
Overall, when comparing the PAH molecules with B,~N-substituted ones, red-shifting of $\Delta\varepsilon$ 
arises due to quasi-degenerate valence MOs characteristic of a diradical-type system, while blue-shifting of $\Delta\varepsilon$ arises due to increase in the $\sigma$ character (decrease in $\pi$ character) of the excitation.
It may be worthwhile to relate this trend to the observation
that 2D BN-sheet is a wide-gap insulator due to lack of extended $\pi$-conjugations\cite{nagashima1995electronic}.


\subsubsection*{E.3. Inter-property correlations}
Rational chemical compound design based on high-throughput DFT computations
often requires multi-property optimization. To this end, we 
explore some of the static ground state properties that are typically computed in a single-point calculation. 
Following $\Delta\varepsilon$, the next key property of interest is the thermodynamic stability of the molecules. For this purpose, we use atomization energy per electron ($E$) as a measure. In addition, ground state dipole moment ($\mu$)---that is routinely computed during single point calculations---contain information about spatial separations of partial charges. In the following, we briefly discuss the correlations between these three properties. 
\begin{figure*}[!htb]
\includegraphics[width=16.5cm]{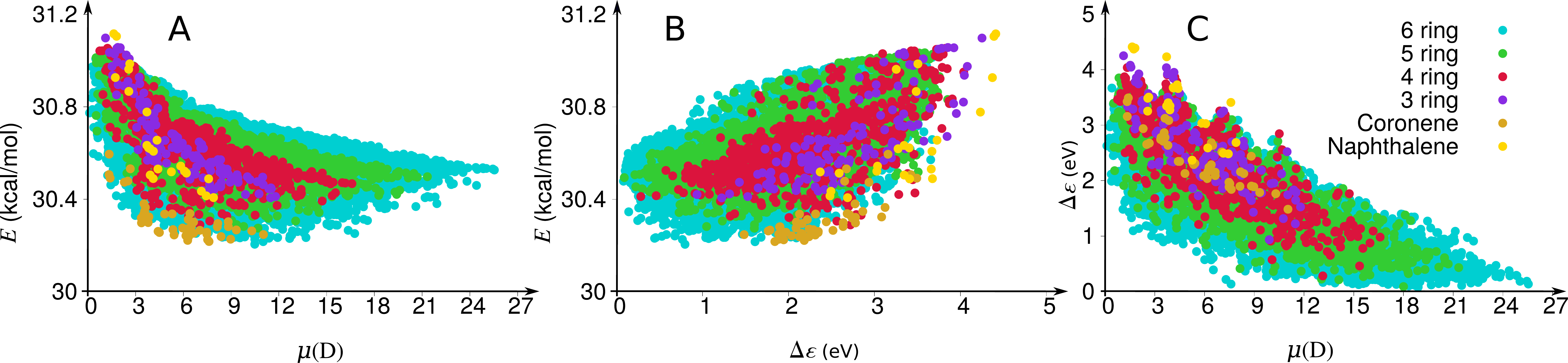}
\caption{Inter-property correlations across 30,797 BN-PAH compounds:
$E$ is the atomization energy per electron in kcal/mol, 
$\mu$ is the dipole moment in debye,
$\Delta\varepsilon$ is the HOMO-LUMO gap in eV.}
\label{DFTgap1}
\end{figure*}

Pairwise inter-property correlations: $E$ vs $\mu$, $E$ vs $\Delta\varepsilon$, $\Delta\varepsilon$ vs $\mu$ are on display in Fig.~\ref{DFTgap1}. For all properties, the
range is largest for the 6-ring compounds; there are 25k such molecules in the 31k set. The spread in the property values decrease gradually with the number of rings.  A noticeable feature in Fig.~\ref{DFTgap1}A is that molecules with large $E$ exhibit small $\mu$. We ascribe this relation to the fact that these molecules have the shortest B-N separations leading to strong bonding. We note in Fig.~\ref{DFTgap1}B that smaller $E$  values typically correlate with smaller $\Delta\varepsilon$ as expected in the case of diradical-type molecules such as those with well-separated B and N centers. Fully-conjugated aromatic molecules show electronic gap in the typical region of about 2-5 eV (see Fig.~\ref{DFTgap1}B) and these molecules are also found to be more stable with larger $E$. Such an interpretation is also supported by the trends shown in Fig.~\ref{DFTgap1}C;
molecules with longer diradical-type bonds show larger $\mu$ and smaller $\Delta\varepsilon$ and {\it vice versa}. These trends are reminiscent of those noted in a previous high-throughput DFT study of 134k small organic molecules\cite{ramakrishnan2014quantum}.

\section{Computational Methods}
For all seventy-seven PAHs listed in the previous section, we have generated Cartesian coordinates using the program Avogadro\cite{hanwell2012avogadro}. With the same program minimum energy  structures were obtained
by employing universal force-field (UFF) parameters\cite{rappe1992uff}. The resulting structures 
were used as templates for combinatorially generating the atomic coordinates of all the B,~N-substituted molecules; permutationally redundant structures were eliminated by comparing the principal moments of inertia. In the case of naphthalene, we have generated all possible molecules where pairs of C atoms were substituted by the isoelectronic B, N atom pairs resulting in 2,285 compounds; while for the larger PAHs comprising of more than two benzene rings, we have restricted the substitution to only a single pair of C atoms which gave rise to  30,774 compounds. These numbers tally perfectly with those from Polya enumeration, as long as the Cartesian coordinates encode the molecular symmetry (see Table~\ref{tab:allcarbons}). For all 33,059 molecules, we have performed geometry optimization and electronic structure calculations at the Kohn--Sham density functional theory (DFT) level using the ORCA (version 4.0.1.2) suite of programs\cite{neese2012orca}. In order to reach a high-degree of quantitative accuracy to model the electronic excitation spectrum one must perform linear-response time-dependent (LR-TD)-DFT calculations preferably based on long-range corrected hybrid DFs and large basis sets. However, in the present study we only wish to provide qualitative and semi-quantitative insights to the stability and HOMO-LUMO gaps of the B,~N-substituted PAHs. For this purpose, we limited our DFT explorations only to the TPSSh\cite{perdew1999accurate,perdew2004meta} hyper-GGA functional---that has been shown to be applicable to model the electronic properties of organic and inorganic molecules \cite{jensen2008bioinorganic,irfan2012quantum,zhang2013cyano,el2014molecular,liu2014novel}---in combination with the split valence basis set def2-SVP \cite{weigend2005balanced}. In all calculations, we have used the 
resolution-of-identity technique to approximate the two-electron Coulomb and exchange integrals (RI-JK approximation) with the corresponding def2/JK \cite{weigend2008hartree} auxiliary basis sets along with {\tt Grid5}-level integration grids to estimate the exchange-correlation energies using numerical quadrature.

\section{Conclusions}
We have applied a combinatorial algorithm to enumerate all possible compounds obtained by substituting a pair of carbon atoms in the smallest seventy-seven poly aromatic hydrocarbons containing 2-6 benzene rings with isoelectronic/isosteric B, N atom pairs. For a hydrocarbon with $N$ carbon atoms, maximal number of compounds are obtained when exactly $N/2$ carbon atoms are substituted by $N$ heteroatom pairs. The grand set of all the resulting compounds is eleven orders of magnitude (7,453,041,547,842=$7.4$ tera) larger than that of the parent hydrocarbons. To facilitate large scale data-mining and discovery of combinatorial trends across the BN-PAH dataset, we have provided consolidated tabulations of molecular distributions according to symmetry, stoichiometry and sites. Furthermore, we show more than one hydrocarbons with same number of carbon atoms and same point group symmetry to lead to distinct yield-patterns revealing a symmetry-controlled selectivity; we have rationalized this effect using the generalized character cycle indices (GCCIs). Our results based on B,~N substitutions are also transferable when using other isovalent heteroatom pairs. 

For a tiny fraction of the 7.4 tera set consisting of 33,059  (33 kilo) representative molecules, we have performed DFT calculations and analyzed structural and electronic features relevant for light-harvesting applications. For the unsubstituted hydrocarbons, we provide qualitative insights into the DFT-predicted properties using Clar's valence electronic structure formulae. Replacing a couple of carbon atoms in the hydrocarbons with a heteroatom pair has been shown to perturbatively modulate all the properties by retaining the essential characteristics of the parent compounds. 
More importantly, our results indicate that combinatorial introduction of B,~N atoms in smallest polycyclic aromatic hydrocarbons gives rise to a library of compounds with HOMO-LUMO gaps spanning the entire solar spectrum,  
with a significant fraction exhibiting HOMO-LUMO gap near
the solar spectral maximum. This prompts us to suggest the suitability of the BN-PAH dataset for various light--harvesting, singlet fission applications\cite{paci2006singlet,greyson2010maximizing,zeng2014seeking} or
to design material with desirable exciton energetics\cite{hill2000charge}.
Designing material exhibiting low exciton binding energies, in order for the absorbed photon to generate maximum output voltage, has been a major theme in studies of organic photovoltaics. The HOMO-LUMO gaps reported in the present work corresponds to the {\it real} gap (a.k.a. transport or fundamental gap), denoted in the relevant literature\cite{bredas2014mind} as $E_g$ or $E_{\rm fund.}$. There have also been studies\cite{bappler2014exciton} addressing how to directly model the so-called {\it optical} gap, $E_{\rm opt.}$, which is the least energy required for the creation of a bound electron-hole pair; this excitation corresponds to the first (narrow) peak in the absorption/photo-luminescence spectra. The difference $E_g-E_{\rm opt.}$ accounts for the exciton binding energy, $E_b$, higher its value lower will be the charge photogeneration efficieny. 
As far as the unsubstituted PAHs are concerned, $E_b$ takes the value of 1 eV for naphthalene and 0.1--0.5 eV for pentacene\cite{lanzani2006photophysics}.  
We hope the BN-PAH dataset along with the presented TPSSh results for $E_g$ could be of use in future high-throughput efforts towards screening materials with small $E_b$.

Recent ventures in comprehensive chemical space design have demonstrated
their pivotal role in accelerating the discovery of novel compounds for a multitude of application domains\cite{shoichet2004virtual,tu2012exploring,balawender2013exploring,reymond2015chemical,ramakrishnan2014quantum,fias2018alchemical}. Such Big Data
efforts also enable rational benchmarking and parameterization of approximate computational chemistry methods in a data-driven fashion\cite{kranz2018generalized,li2018density}. For more elaborate design studies based on the BN-PAH compound library the results provided in the present work may be
considered as a baseline. 

\section{Supplementary Material}
For all seventy-seven PAH,
complete pattern inventory for all possible B, N substitutions are
collected. For seventy-seven PAH and 33,059 BN-PAH molecules, TPSSh/def2-SVP/RI-JK-level equilibrium structures and various electronic properties are also collected. 

\begin{acknowledgements}
We gratefully acknowledge Prof. Gunnar Brinkmann for providing the {\tt CaGe} program, Prof. Ranjan Das and Dr. Vamsee Voora for useful discussions. PK is grateful to TIFR for Visiting Students’ Research Programme (VSRP) and junior research fellowships. RR and SC thank TIFR for financial support. All calculations have been performed using the {\tt helios} computer cluster which is an integral part of the {\tt MolDis} Big Data facility, TIFR Hyderabad ({\tt https://moldis.tifrh.res.in/}). 
\end{acknowledgements}
\section*{References}
\bibliography{aipsamp}

\end{document}